\documentclass[aps, prl, singlecolumn, reprint, groupedaddress]{revtex4-1}

\usepackage{graphicx}% Include figure files
\usepackage{dcolumn}% Align table columns on decimal point
\usepackage{bm}% bold math
\usepackage{color}

\begin{document}

\preprint{AIP/123-QED}

%\title[Sample Title]{Sample Title:\\with Forced Linebreak\footnote{Error!}}% Force line breaks with \\
%\thanks{Footnote to title of article.}

\title[]{Josephson parametric converter saturation and higher order effects}
%\thanks{Footnote to title of article.}

\author{G. Liu, T.-C. Chien, X. Cao, O. Lanes, E. Alpern, D. Pekker,  and M. Hatridge}
\email{hatridge@pitt.edu.}
\affiliation{Department of Physics and Astronomy, University of Pittsburgh, Pittsburgh, Pennsylvania 15260, USA}

\date{\today}% It is always \today, today,

\begin{abstract}

Microwave parametric amplifiers based on Josephson junctions have become indispensable components of many quantum information experiments.  One key limitation which has not been well predicted by  theory is the gain saturation behavior which limits the amplifier's ability to process large amplitude signals. The typical explanation for this behavior in phase-preserving amplifiers based on three-wave mixing, such as the Josephson Parametric Converter (JPC), is pump depletion, in which the consumption of pump photons to produce amplification results in a reduction in gain.  However, in this work we present experimental data and theoretical calculations showing that the fourth-order Kerr nonlinearities inherent in Josephson junctions are the dominant factor.  The Kerr-based theory has the unusual property of causing saturation to both lower and higher gains, depending on bias conditions.  This work presents a new methodology for optimizing device performance in the presence of Kerr nonlinearities while retaining device tunability, and points to the necessity of controlling higher-order Hamiltonian terms to make further improvements in parametric devices.

\end{abstract}

\maketitle

Quantum-limited amplification is a vital tool in quantum information processing.  At microwave frequency, such amplifiers enable high-fidelity measurement of quantum bits\cite{Vijay2011,Hatridge2013}, nano-mechanical resonators\cite{Teufel2011} and flying states of light\cite{Eichler2012}.  The amplifiers  are typically built from microwave resonators containing one or more superconducting Josephson junctions, which provide the essential non-linear Hamiltonian terms\cite{Yamamoto2008,Vijay2009,Roy2016}. The strength of the non-linear coupling between the device's modes are controlled via an external microwave pump, which in turn sets the amplifiers' gain and center frequency, hence their collective description as Josephson Parametric Amplifiers (JPAs)\cite{Lehnert20008,Lehnert2007}. 

A JPA's utility is determined by several parameters.  These include its quantum efficiency, which describes the noise added during amplification\cite{Caves1982},  tunability to match the signal frequency of interest, instantaneous bandwidth, and the ability to process large amplitude signals.  This last parameter is typically referred to as the saturation power, or more precisely as $P_{\textrm{-1dB}}$, the input power at which the gain falls by 1~dB from its small signal value. Conventionally, saturation in JPAs is attributed to depletion of photons from the microwave pump tone\cite{Abdo2013,Eichler2014}.  The pump both controls the amplifier gain and serves as the power source for photons created in the amplification process, resulting in a monotonic decrease in gain with increasing signal power.

However, pump depletion has, in almost all cases, failed to give an accurate description of experimental device performance. In this letter we show that,  instead, Kerr nonlinearities inherent to Josephson junctions are the dominant factor that limits device saturation power.  Our results give good qualitative agreement between a theory which completely neglects pump depletion and experimental data for phase preserving amplification in the Josephson Parametric Converter (JPC)\cite{Bergeal2010,Bergeal2010a}.   We find that for typical device parameters, the Kerr terms of the Hamiltonian cause the system to dispersively shift away from its bias point before the effects of pump-depletion become relevant.

Given this new understanding, we present a methodology for optimizing device performance in the presence of Kerr nonlinearities while retaining device tunability.  Although in this paper we specifically study amplifiers based on three wave mixing with the Josephson Ring Modulator (JRM)\cite{Abdo2013}, this effect will be equally prominent in three-wave mixing devices based on SQUIDS  or other multi-junction circuits with similar-amplitude Kerr terms\cite{Lecocq2016}.  We note that a related effect has been studied theoretically for the case of single junction four-wave mixing based amplifiers \cite{Bogdan2015}.
\begin{figure}[h]
	\includegraphics[scale = 0.9]{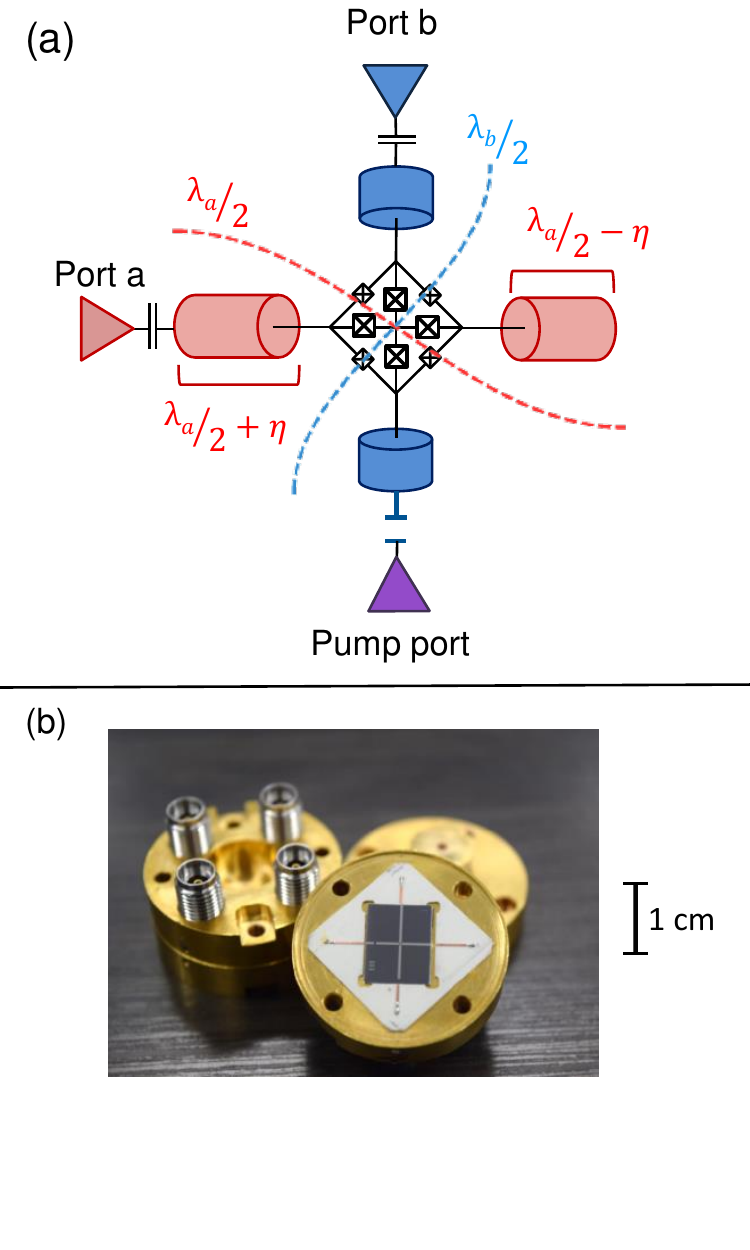}
	\caption{\label{fig:jpc} {\small \textbf{(a)} Schematic of a single-ended Josephson Parametric Converter (JPC) circuit. The device consists of two $\lambda /2$ resonators which meet at a central ring of Josephson junctions, the JRM.  The horizontal (red) mode is labeled $a$, the vertical (blue) mode $b$, and there is a third, common mode $c$ of the two arms. Resonant modes $a$ and $b$ are each strongly coupled to a single microwave port, the pump is weakly coupled via the pump port.  \textbf{(b)} Image of assembled JPC.   The $a$ and $b$ modes and pump port are each accessed through individual SMA connectors.}}
\end{figure}

The JPC realizes non-degenerate three-wave mixing with a ring of four nominally identical Josephson junctions (the JRM), placed at the intersection of two $ \lambda/2$ resonators (see Fig.~1a). The horizontal mode is referred to as the idler or $a$-mode, while the vertical mode is the signal, or $b$-mode. There is a third, common mode, $c$, consisting of a joint excitation of the horizontal and vertical spatial modes. The signal and idler mode are each strongly coupled to a single microwave port accessible through transmission lines with decay rates $\kappa_{a,b}$ while the pump tone is coupled to the $c$ mode via a weakly coupled pump port. The device tunability is enhanced by the addition of four interior junctions, which are much larger than the outer junctions that produce the three wave mixing and as such are treated as linear inductors\cite{Roch2012}. Up to third order in creation/annihilation operators, the Hamiltonian of the JPC in the rotating wave approximation can be written as\cite{Bergeal2010}, 
\begin{equation}
	\frac{H_{\rm{JPC}}}{\hbar} = \omega_a a^\dag a + \omega_b b^\dag b + \omega_c c^\dag c + g (a^\dag b^\dag c + a b c^\dag)
	\label{eq:JPCH}
\end{equation}
where $a$, $b$ and $c$ are annihilation operators of the three modes of the JPC, and $g$ is the flux-dependent three-wave coupling strength.  Gain is achieved by applying a strong microwave drive  to  spatial mode $c$ at the frequency $\omega_p \simeq \omega_a + \omega_b$. If this is strongly detuned from any $c$-mode resonance, the pump is said to be ``stiff'', meaning $c$ can be replaced with its average value. In this letter, we calculate the average response of the amplifier using semi-classical Langevin equations derived from the circuit Hamiltonian, together with the modes' coupling rates to the microwave environment (see supplement Sec.~I). 

%The 3-wave mixing term $g$ is flux dependent, with zero amplitude at integer flux quanta applied to the JRM. Thus, the device must be flux-biased away from these points in order to operate.  Combined with the flux dependence of the mode frequencies $\omega_{a,b}$, varying the applied flux allows amplification over a wide range of frequencies. 
The flux dependence of the mode frequencies $\omega_{a,b}$ allows amplification over a wide range of frequencies by varying the flux applied to the JRM. At a fixed flux, the amplifier can be further tuned over a narrower range of frequencies, roughly corresponding to mode bandwidths $\kappa_{a,b}$, by varying the pump frequency away from the sum frequency by $\epsilon$, so that $\omega_p = \omega_a + \omega_b+\epsilon$. Note that for each pump frequency there is a unique peak gain frequency which depends on both the pump detuning and the mode bandwidths. For $\kappa_{a,b}=\kappa \mp \Delta\kappa/2$ the peak gain frequency (for mode $a$) can be written to first order in $\epsilon$ and $\Delta\kappa$ as $\omega_{\tiny{\textrm{max G}}}=\omega_a+\left(\frac{1}{2}-\frac{\kappa \Delta\kappa/2}{4 g^2 \left<c\right>^2 + \kappa^2+\left(\Delta\kappa/2\right)^2}\right) \epsilon$ (see supplement Sec.~I).   

In the JPC Hamiltonian, the 4th order (Kerr) nonlinearity, which is typically neglected in Eq.~\ref{eq:JPCH}, are
\begin{equation}
\frac{H_{\rm{Kerr}}}{\hbar}  = - \sum_{m=a}^c \sum_{n=m}^c K_{mn} a_m^\dag a_m a^\dag_n a_n\\
\label{eq:H4}
\end{equation}
where $a_{i}=\left(a,b,c\right)$ and $K_{mn}$ are the Kerr amplitudes (see supplement Sec.~II). Given the stiffness of the pumped $c$ mode, the $K_{cc}$ term is a constant for a given set of pump conditions and can be neglected, leaving five terms to be considered. Of these, the $K_{ac}$ and $K_{bc}$ are simplified by the stiff pump approximation to be pump-dependent Stark shifts of the $a$ and $b$ modes.  Their effect is visible even at very low signal power as they shift the optimal pump frequency (defined as the frequency requiring minimum pump power for achieving a given gain for very low signal powers) to be smaller than the sum frequency of the $a$ and $b$ modes.

\begin{figure*}[t]
	\includegraphics[scale = 1]{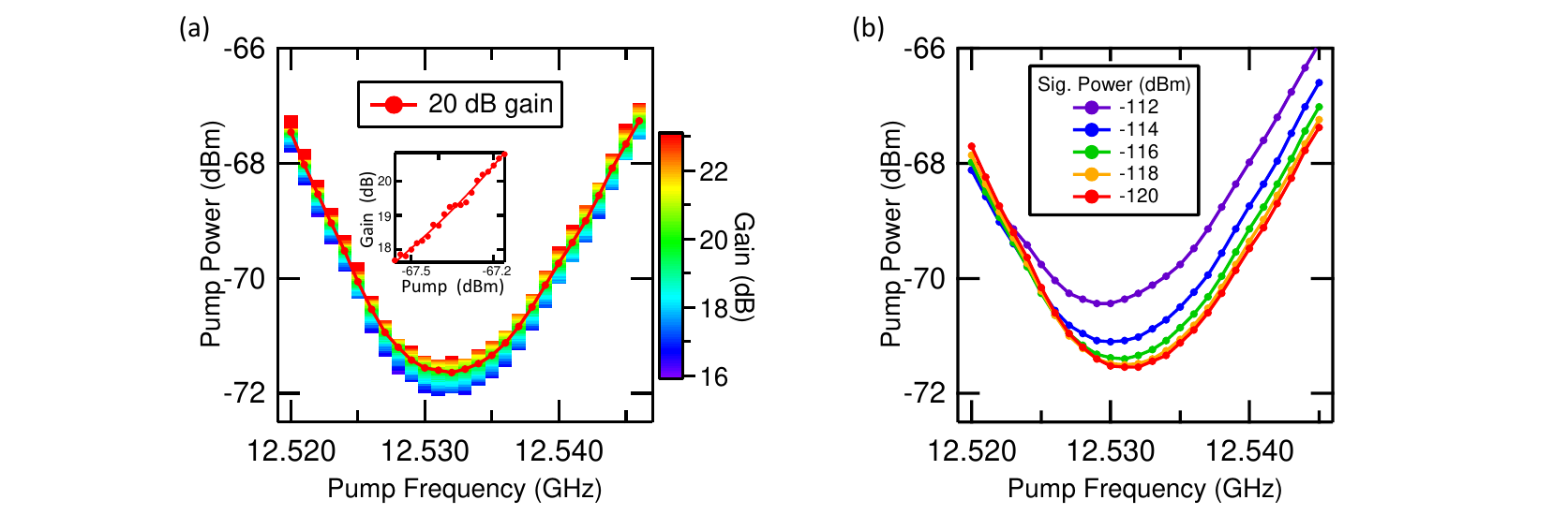}
	\caption{\label{fig:20dB_pfreq_ppower} {\small \textbf{(a)} Maximum gain vs. pump frequency and power for -140~dBm signal power. Each pixel represents the fitted maximum gain for a pump power/frequency combination.  The red line connects all the 20~dB points obtained from fitting gain data at each pump frequency vs pump power (see inset for example fit curve).  \textbf{(b)} Experimental $G=20$~dB points versus pump frequency and power for varying signal powers.}}
\end{figure*}

The final three terms, $K_{aa}$, $K_{bb}$ and $K_{ab}$ grow with signal power and give a further nonlinear contribution to the amplifier response.  Their contributions are largely indistinguishable, as the process of phase-preserving amplification results in very tightly correlated $a$ and $b$ mode populations\cite{Flurin2012}.  The increasing signal amplitude causes the coupled modes to shift to lower resonant frequencies and increased nonlinear response as a function of signal power, reminiscent of the behavior of single mode Duffing oscillators \cite{Vijay2008,Kovacic2011}.  In general, we can only solve these equations numerically: to gain an intuitive picture of the system's behavior as a function of signal power we calculate the pump power which gives fixed power gain  $G$ at different pump frequency for varying signal power (see supplement Sec. II).  The results (supplemental Fig. S2) show an apparent shift of the curve to lower frequencies and higher pump powers as the coupled modes dispersively shift away from the pump tone.

All experimental data was taken from a single-ended JPC shown schematically in Fig.~\ref{fig:jpc}a. The need for hybrids to couple symmetrically to the $a$ and $b$ modes has been eliminated. The resulting asymmetry between the two ends of the resonators shifts the current anti-nodes away from the JRM and results in leakage between modes $a$ and $b$.  We correct this effect by introducing an offsetting asymmetry ($\eta$) in the length of the two arms of each resonator as indicated in the figure. Figure \ref{fig:jpc}b shows an image of the assembled JPC. The device is fabricated using double-angle aluminum deposition of Josephson junctions and resonator on silicon together with a 1.5~$\mu$m silver ground plane on the reverse side. The critical current for the outer junctions is 1.78~$\mu$A, and for the inner junctions is 5.34~$\mu$A. The modes $a$, $b$ and $c$ are each accessed through individual SMA connectors.

For all data, an external DC magnetic flux, $\Phi_{ext} = 1.2~\Phi_0$, was applied, where $\Phi_0 = h/{2e} $ is the magnetic flux quantum and we define the flux as applied to the full JRM (which with its four loops is periodic with $4 \Phi_0$ total applied flux). At this flux, the resonant frequencies of mode $a$ and  $b$ are $\omega_a /2\pi = 5.0847$~GHz and $\omega_b/2 \pi = 7.4471$~GHz, and the line-widths are $\kappa_a /2\pi= 20.27$ MHz and $\kappa_b/2 \pi = 62.17$ MHz. We first identified the combination of pump powers and frequencies yielding $G=20$~dB, as shown in Fig.~2a.  For each pixel, a pump power and frequency combination were applied to the pump port, and the small-signal response for $P_{sig}=-140$~dBm was recorded.  Each curve was first fitted to identify the maximum gain and associated signal frequency. We found that the most accurate bias conditions were identified by subsequently fitting all peak-gain points at a given pump frequency to the expect response of $G$ vs. $P_p$, as shown in the inset.  

Next, we evaluated the influence of increasing signal power by repeating this protocol for increasing signal powers, as shown in Fig.~2b.  As the signal power increases, the amplifier response shifts to lower frequency in excellent qualitative agreement with calculated results (see supplement Fig.~S2a), including the asymmetry between positive and negative detunings. For positive detunings the modes shift away from the bias point, thus higher pump power is required to maintain $20$~dB gain.  For negative detunings, the situation is at first reversed as the modes move closer, resulting in an initial shift to \textit{higher} gain before they, too, fall as the modes continue to shift with increasing signal power. 

This anomalous behavior requires us to modify how we evaluate saturation, otherwise we may assign very high saturation powers to an amplifier whose response is extremely nonlinear.  A more symmetric limit of $P_{\pm\textrm{1dB}}$, defined as the power at which the gain first deviates in either direction by 1~dB from its small signal value, will give a much fairer comparison of different bias conditions.  

The amplifier's saturation behavior was measured as shown in Figure~3. For each pump frequency we recorded gain vs. signal power while using the pump power and signal frequency determined in Fig.~2b. The full data set is shown in Fig.~3b; for clarity, representative curves are plotted separately in Fig.~3a.  The calculated saturation curves using extracted device parameters (see supplement Sec.~II) are plotted in Fig.~3c.  For both data and calculation, the gain initially increases with signal power at negative detuning before finally falling, and for positive detunings the gain monotonically decreases.  The $\pm1$~dB saturation values are indicated by red triangles and blue diamonds, respectively.  
\begin{figure*}[t]
	\includegraphics[scale = 1]{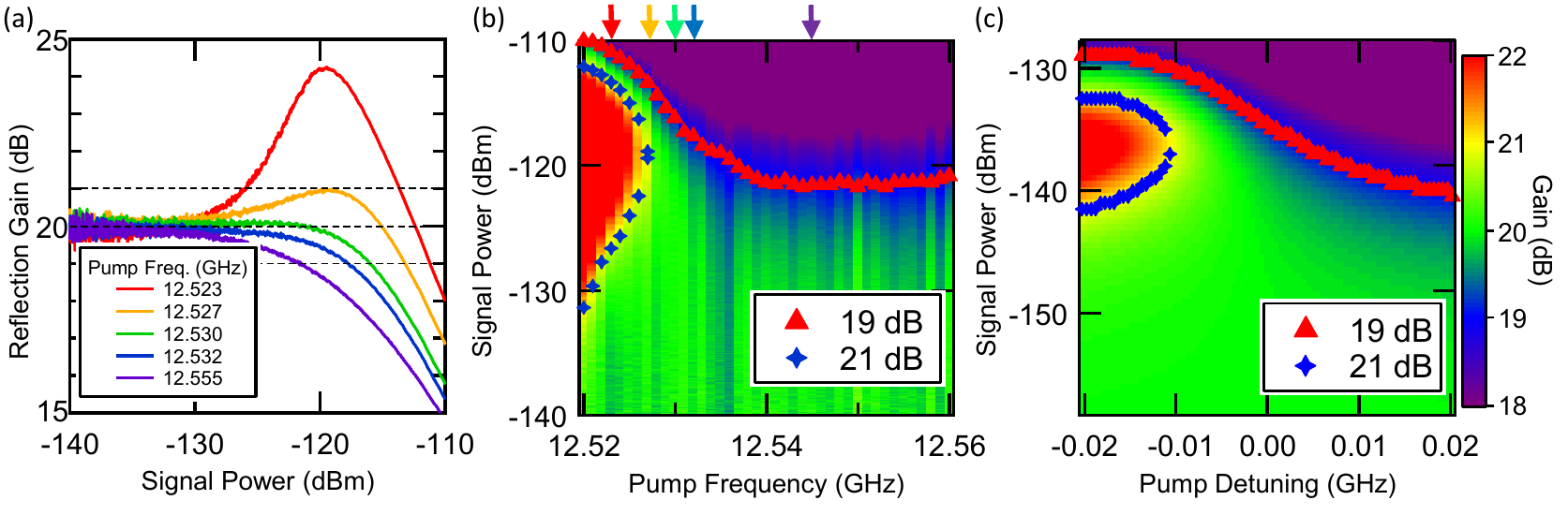}
	\caption{\label{fig:DR_exp_vs_thry} {\small \textbf{(a)} Measured reflection gain of the JPC vs. signal power at selected pump frequencies, showing the variation in saturation response vs pump frequency from the resonant condition.  \textbf{(b)} Measured gain vs. signal power and pump frequency for $20$~dB bias conditions identified in Fig.~2b. \textbf{(c)} Calculated theoretical gain at different signal powers and pump frequencies.  In both (b) and (c) the  saturation values are indicated as red triangles ($-1$~dB) and blue diamonds ($+1$~dB).}}
\end{figure*}

At positive detunings the $P_{\textrm{-1dB}}$ limit is reached first, in both theory and experiment, eventually leveling off at a value 5-10~dB lower (Fig.~3a blue and purple data) than the optimal monotonically decreasing gain point(Fig.~3a in green), which is found very near the small-signal resonant condition.  For negative detunings, the gain rise phenomenon becomes increasingly severe (Fig.~3a in red), eventually resulting in unstable/hysteretic gain conditions (not shown).  However, for modestly negative detuning, the gain rise phenomenon can act to enhance the saturation power.  Thus, we identify an alternate optimum bias condition (Fig.~3a in orange) which rises to just less than $+ 1$~dB before falling.  Taken together, these factors can result in amplifier performance that varies by well over $10$~dB if the amplifier is biased without knowledge of the Kerr effect.  As most amplifiers operate over a modest range of bandwidths (10's-100's of MHz) and critical currents (few $\mu$A), these behaviors should be visible in all devices, and are, in fact, visible in previously published data (for example in Ref.~9).  
\begin{figure}[h!]
	\includegraphics[scale=0.8]{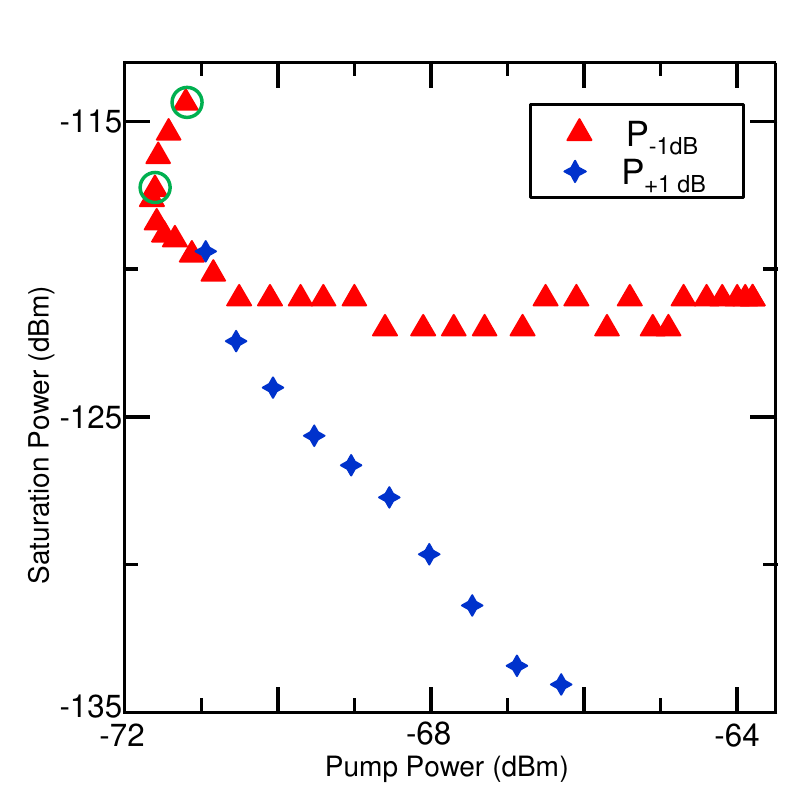}
	\caption{\label{fig:PumpPower_PumpFreq_shift} {\small $P_{\pm\textrm{1dB}}$ point for different pump power. The $P_{\textrm{-1dB}}$  data shows the saturation of the dynamic range on the positive detuning side, while the $P_{\textrm{+1dB}}$ data shows that the dynamic range keeps decreasing as the pump power increases. Green circles indicate the two optimal points.}}
\end{figure}

Gain saturation is summarized in Figure~4, colored by which limit ($\pm 1$~dB) is reached first. For positive detuning the $P_{\textrm{-1dB}}$ (red triangles) limit is relevant and falls to a static value even as $P_{p}$ continues to climb for increased detuning.  For negative detuning, the $P_{\textrm{+1dB}}$ (blue diamonds) behavior is limiting and falls steadily with increasing $P_p$ as the amplifier's response becomes increasingly distorted. The two optimum points are circled, and are both near the lowest pump powers, in direct contradiction to the expectations of pump depletion theory that bias conditions requiring stronger pump should yield higher saturation powers.  

We note that although this result suggests that the JPC possesses only one best bias point for each bias flux, by jointly varying the pump frequency and flux the device should be no less tunable. In fact, our result suggests that the device can be readily tuned by jointly varying flux and pump frequency to minimize the pump power required for a given signal frequency. Finally, we add a caution that this picture can be severely disrupted by variation in the impedance presented by the microwave lines connected to the device modes, unless great care is taken to minimize reflections and mismatches in the microwave cabling. In our experiment this is the dominant source of disagreement between theory and experiment, as the device bandwidth is observed to vary significantly for the range of frequencies at which we recorded gain data.  However, at all bias points the overall behavior of Kerr-based shifts to lower frequencies dominated the device performance and allowed us to identify optimal bias conditions.
	
In conclusion, we have developed a theoretical treatment which neglects the dynamics and depletion of the microwave pump and focuses on the fourth-order Kerr terms as the source of amplifier saturation.  Our data and calculations are in excellent qualitative agreement, and we identify a new paradigm for operating three-wave parametric amplifiers in the presence of Kerr nonlinearity. Our results also have vital implications for recent efforts to build multi-parametric Josephson devices, such as directional amplifiers and circulators \cite{Ranzani2014a, Metelmann2015, Sliwa2015a, Lecocq2016}. These devices require the delicate matching of several parametric processes spanning multiple modes, providing a very difficult challenge to tune up if the modes themselves move with changing pump conditions.  The fourth-order theory can be readily extended to these devices, and will provide much needed insight into both bias conditions and saturation behavior.  

However, to make substantial improvements in device performance we must eliminate unwanted higher-order terms through Hamiltonian design. We calculate that  a reduction in Kerr term amplitude translates to an equal increase in saturation power until either pump depletion or the sixth-order terms dominate the device response.  There has been a very recent effort to achieve such a reduction by using an asymmetric flux-biased Josephson circuit (the so-called `SNAIL') to replace the individual junctions in the JRM~\cite{Frattini2017}. 

\begin{acknowledgments}
The authors wish to acknowledge work by E. Brindock and A. Rowden on control software. This manuscript is based upon work supported in part by the U.S. Army Research Office under grant number W911NF-15-1-0397.
\end{acknowledgments}

\bibliography{bibwkmsmtvol2a}% Produces the bibliography via BibTeX.
\end{document}

%\title[Sample Title]{Sample Title:\\with Forced Linebreak\footnote{Error!}}% Force line breaks with \\
%\thanks{Footnote to title of article.}

\title[]{Supplementary Material for `Josephson parametric converter saturation and higher order effects'}

\author{G. Liu, T.-C. Chien, X. Cao, O. Lanes, E. Alpern, D. Pekker,  and M. Hatridge}
\email{hatridge@pitt.edu.}
\affiliation{Department of Physics and Astronomy, University of Pittsburgh, Pittsburgh, Pennsylvania 15260, USA}
%\thanks{Footnote to title of article.}


\maketitle

\section{I. Semi-classical solution for the Josephson Parametric Converter: third order}

The Josephson Parameteric Converter (JPC) consists of a Josephson Ring Modulator (JRM)  coupled to  microwave resonators. The JRM, which in its simplest form is comprised of four identical Josephson junctions arranged in a superconducting loop, has three different spatial modes ($a$, $b$ and $c$), each of which couples to an external microwave mode with matching spatial configuration.  The Hamiltonian of the JRM can be written as 
\begin{equation}
\begin{split}
H_{JRM} = & -4 E_J\cos \left({\frac{\Phi_{ext}}{4\varphi _0}}\right)\cos\left({\frac{\Phi_a}{2\varphi _0}}\right)\cos\left({\frac{\Phi_b}{2\varphi _0}}\right)\cos\left({\frac{\Phi_c}{\varphi _0}}\right)\\ {}{} & -4 E_J\sin\left({\frac{\Phi_{ext}}{4\varphi _0}}\right)\sin\left({\frac{\Phi_a}{2\varphi _0}}\right)\sin\left({\frac{\Phi_b}{2\varphi _0}}\right)\sin\left({\frac{\Phi_c}{\varphi _0}}\right),
\end{split}
\label{eq:JRM}
\end{equation}
where $E_J$ is Josephson junction energy, $\varphi_0=\hbar/2e$ is the reduced flux quantum, $\Phi_i$ is the flux for the $i$-th spatial mode, and $\Phi_{ext}$ is the external flux applied to the JRM loop . Typically $\Phi_{a,b,c}\ll 2\pi \varphi_0$, so the cosine and sine terms can be expanded as power series. By ignoring terms higher than third order in $\Phi_i$, the Hamiltonian of the JRM can be written as 
\begin{equation}
H_{JRM}=\frac{E_J}{2{\varphi_0}^2}\cos\left(\frac{\Phi_{ext}}{4\varphi _0}\right)\left({\Phi_a}^2+{\Phi_b}^2+{\Phi_c}^2\right)- \frac{E_J}{2{\varphi_0}^2}\sin\left(\frac{\Phi_{ext}}{4\varphi _0}\right){\Phi_a}{\Phi_b}{\Phi_c}
\end{equation}
The second term on the right hand side gives the three-wave coupling between the spatial modes of the JRM which leads to amplification and frequency conversion of microwave photons. Next we add the inductance and capacitance of the microwave resonators, which we approximate on resonance as LC oscillators (see \cite{Bergeal2010}).  Their contribution to the total JPC energy is expressed as 

\begin{equation}
H_{res}=\frac{{Q_a}^2}{2C_a}+\frac{{\Phi_a}^2}{2L_a}+\frac{{Q_b}^2}{2C_b}+\frac{{\Phi_b}^2}{2L_b}+\frac{{Q_c}^2}{2C_c}+\frac{{\Phi_c}^2}{2L_c}.
\end{equation}

For the $\lambda/4$ segments of transmission line used in our circuit, the effective inductances are in series with the respective JRM spatial modes.  To express the total Hamiltonian of the JPC, we introduce a new set of conjugate canonical variables, $\tilde{\Phi}_i$, $\tilde{Q}_i$ which we calculate using the concept of a series participation ratio  ($p_i$) as in \cite{Schackert2013} which expresses the fraction of the total mode energy resident in a given JRM mode. The JPC's Hamiltonian can then be expressed as
\begin{equation}
\begin{split}
H_{JPC}= & \frac{{\tilde{Q}_a}^2}{2C_a}+\frac{{\tilde{\Phi}_a}^2}{2L^\prime_a}+\frac{{\tilde{Q}_b}^2}{2C_b}+\frac{{\tilde{\Phi}_b}^2}{2L^\prime_b}+\frac{{\tilde{Q}_c}^2}{2C_c}+\frac{{\tilde{\Phi}_c}^2}{2L^\prime_c} \\
{}{} & - \frac{E_J}{2{\varphi_0}^3} \sin\left(\frac{\Phi_{ext}}{4\varphi _0}\right)p_{a}p_{b}p_{c}{\tilde{\Phi}_a}{\tilde{\Phi}_b}{\tilde{\Phi}_c}
\end{split}
\end{equation}
where 
\begin{equation}
{L^\prime}_{a,b,c}=L_J\left(\Phi_{ext}\right)+L_{a,b,c}\;,\;L_J\left(\Phi_{ext}\right)=\frac{{\varphi_0}^2}{E_J \cos(\frac{\Phi_{ext}}{4\varphi_0})}\;,\;
\textrm{and }p_{i}=\frac{L_J\left(\Phi_{ext}\right)}{L_J\left(\Phi_{ext}\right)+L_i},\; i \in \{a,b,c\}
\end{equation}

Canonical variables can be transformed into creation and annihilation operators through the following relation
\begin{equation}
\tilde{\Phi}_j=\sqrt{\frac{\hbar Z_j}{2}}(j+j^\dagger)\textrm{ };\textrm{ }
\tilde{Q}_j=i \sqrt{\frac{\hbar}{2 Z_j}}(j-j^\dagger)\textrm{ };\textrm{ }
Z_j=\sqrt{\frac{L_j+L_J}{C_j}}\textrm{ };\textrm{ }
j \in \{a,b,c\}
\end{equation}

With a strong pump $\omega_p\equiv\omega_1+\omega_2\simeq\omega_a+\omega_b$ applied to the $c$-mode, and under the rotating wave approximation (RWA) we arrive at 
\begin{equation}
\frac{H_{JPC}}{\hbar}=\omega_{a} a^\dagger a+\omega_{b} b^\dagger b+\omega_{c} c^\dagger c+g\left(a^\dagger b^\dagger c+ab c^\dagger\right)
\label{eq:JPCH3}
\end{equation}
where 
\begin{equation*}
g=-\frac{ p_{a}p_{b}p_{c}E_J\sqrt{\hbar}}{2\sqrt{2}{\varphi_0}^3} \sin\left(\frac{\Phi_{ext}}{4\varphi _0}\right)\left(\frac{L^\prime_a}{C_a}\frac{L^\prime_b}{C_b}\frac{L^\prime_c}{C_c}\right)^{1/4}
\end{equation*}
A schematic of the operation of the JPC is given in Fig.~S1.  Signals incident on port $a,b$ at angular frequency $\omega_{1,2} \simeq \omega_{a,b}$ will be amplified both at the same frequency with voltage gain $\sqrt{G}$ in reflection and transmitted with amplitude $\sqrt{G-1}$ to port $b,a$ at angular frequency $\omega_{2,1}=\omega_p-\omega_{1,2}$ with nonreciprocal phase-shift $\pm\phi_p$ set by the pump tone.

\begin{figure*}[h]
	\includegraphics{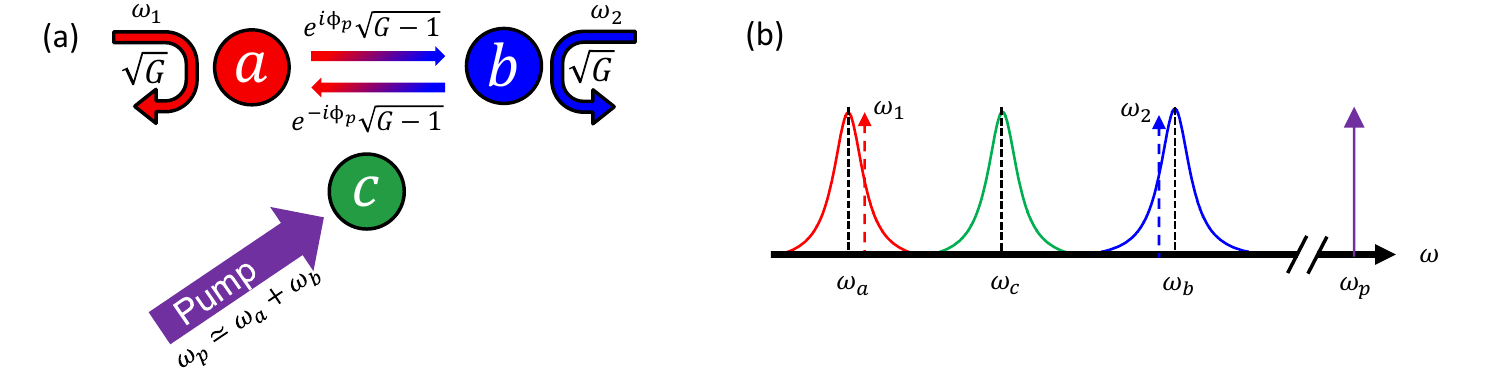}
	\caption{\label{fig:jpc} (a) Diagram of phase-preserving amplification of inputs to modes $a$ and $b$. (b) Schematic of JPC modes and pump tone in frequency space. The red and blue Lorentzian line shapes represent mode $a$ and $b$ of the JPC. The purple and blue arrow represent the pump tone and input signal to mode $b$. The dashed red arrow corresponds to signal generated into mode $a$ through the parametric gain process. }
\end{figure*}

In this work, we assume a perfectly `stiff' pump tone applied to port $c$, which is to say that the pump is sufficiently intense  and detuned from any $c$-mode resonance that the change in pump amplitude is negligible despite the fact that pump-photons are being converted to signal and idler photons to produce amplification. Then in Eq.~\ref{eq:JPCH3} $c$ can be replaced by it's classical average value $\langle c \rangle$. Next, we treat the effect of the mode's coupling to the external environment by constructing semi-classical Langevin equations of the JPC in frequency domain as in \cite{Clerk2010, Roy2016}.  We specialize to the case of no signal incident on the $b$-mode ($b_{in}=0$) and express the reflection and transmission voltage gain as $\alpha=a^{out}/a^{in}$ and $\beta={b^{out}}^\dagger/a^{in}$ respectively.  This yields
\begin{equation}
\left(\frac{\kappa_a}{2}-i \Delta\right) \frac{\left(1+\alpha\right)}{\sqrt{\kappa_a}}=-i\frac{g\langle c \rangle}{\sqrt{\kappa_b}} \beta + \sqrt{\kappa_a}
\end{equation}
\begin{equation}
\left(\frac{\kappa_b}{2}-i \left(-\Delta+\epsilon\right)\right) \frac{\beta^*}{\sqrt{\kappa_b}}=-i\frac{g\langle c \rangle}{\sqrt{\kappa_a}}\left(1+\alpha^*\right),
\end{equation}
where $\Delta=\omega_1-\omega_a=\omega_b-\omega_2+\epsilon$, and $\epsilon=\omega_p-\omega_a-\omega_b$. Notice that $\alpha$ and $\beta$ are complex variables. These coupled equations are solved analytically for the reflected power gain: 
\begin{equation}
{|\alpha|}^2=\frac
{16g^4 {n_p}^2+8g^2{n_p}\left(4\Delta\left(\Delta-\epsilon\right)+{\kappa_a}{\kappa_b}\right)+\left(4\Delta^2+{\kappa_a}^2\right)\left(4{\left(\Delta-\epsilon\right)}^2+{\kappa_b}^2\right)}
{16g^4 {n_p}^2+8g^2{n_p}\left(4\Delta\left(\Delta-\epsilon\right)-{\kappa_a}{\kappa_b}\right)+\left(4\Delta^2+{\kappa_a}^2\right)\left(4{\left(\Delta-\epsilon\right)}^2+{\kappa_b}^2\right)}
\label{eq:Refgain}
\end{equation}
where $n_p=\langle c^\dagger c \rangle$. 

We obtain the signal frequency where peak gain occurs for a given pump frequency and amplitude by solving for the frequency at which $\frac{\partial|\alpha|^2}{\partial\Delta}=0$.  The behavior of the maximum gain frequency depends crucially on the difference between $\kappa_a$ and $\kappa_b$, so we make the substitutions $\kappa_{a,b}=\kappa \mp \frac{\Delta \kappa}{2}$.  In general, there is a non-linear dependence on the pump detuning $\epsilon$, for small detuning we can write to first order in $\epsilon$ 
\begin{equation}
\Delta_{\tiny{\textrm{max G}}} = \left(\frac{1}{2}-\frac{\kappa \left(\Delta \kappa /2\right)}{4g^2 n_p+\kappa^2+\left(\Delta \kappa /2\right)^2}\right) \epsilon+{O\left[\epsilon\right]}^3
\end{equation}
We note that for $\Delta\kappa=0$ this expression greatly simplifies and the peak gain frequency does not vary vs. pump power, and hence gain.  However, for dissimilar resonator loss rates, even in this third order treatment the peak frequency will shift vs. pump power if the pump frequency differs from the modes' sum frequency.  We experimentally fit each gain curve for a unique peak gain amplitude and $\Delta_{\tiny{\textrm{max G}}}$,  then substitute $\Delta_{\tiny{\textrm{max G}}}$  into the expression for $|\alpha|^2$ we obtain the familiar result \cite{Bergeal2010}
\begin{equation}
G = |\alpha|^2=\left(\frac{1+\frac{P_p}{P_c}}{1-\frac{P_p}{P_c}}\right)^2,
\label{eq:Gsimp}
\end{equation}
where the pump power $P_p\propto n_p$ and $P_c$ is the critical power at which gain diverges.  We use this equation to fit the $G$ vs. $P_{p}$ for all data in Fig.~2b and c.

\section{II. Semi-classical solution for JPC: fourth order}

Next, we extend our treatment to include the fourth order terms in the expansion of Eq.~\ref{eq:JRM}. The Hamiltonian of the JPC becomes
\begin{multline}
\frac{H_{JPC}}{\hbar}=
{\omega_a}^\prime a^\dagger a+{\omega_b}^\prime b^\dagger b+{\omega_c}^\prime c^\dagger c+g \left(a^\dagger b^\dagger c + a b c^\dagger\right)\\
-\frac{1}{2}\text{K}_{aa}a^\dagger a^\dagger a  a-
\frac{1}{2}\text{K}_{bb}b^\dagger b^\dagger b  b-
8\text{K}_{cc}c^\dagger c^\dagger c  c-2\text{K}_{ab}a^\dagger a b^\dagger b-8\text{K}_{ac}a^\dagger a c^\dagger c-8\text{K}_{bc}b^\dagger b c^\dagger c
\end{multline}
in which the mode frequencies shift slightly to
\begin{equation*}
{\omega_a}^\prime=\omega_a-\frac{1}{2}\text{K}_{aa}-\text{K}_{ab}-4\text{K}_{ac},\;
{\omega_b}^\prime=\omega_b-\frac{1}{2}\text{K}_{bb}-\text{K}_{ab}-4\text{K}_{bc}\;\text{and}\;
{\omega_c}^\prime=\omega_c-8\text{K}_{cc}-4\text{K}_{ac}-4\text{K}_{bc}
\end{equation*}
and 
\begin{equation}
\text{K}_{ii}=\frac{\hbar E_J}{32 \varphi_0^4}\cos\left({\frac{\Phi_{ext}}{4 \varphi_0}}\right) {p_{i}}^2 \frac{L_i}{C_i}
\end{equation}
\begin{equation}
\text{K}_{ij}=\sqrt{\text{K}_{ii}\text{K}_{jj}}
\end{equation}
defined as the `self' and `cross' Kerr terms for the system, respectively.
As before, we assume a `stiff' pump and no input signal on mode $b$, yielding the modified semi-classical quantum Langevin equations for the JPC:
\begin{equation}
\left[\frac{\kappa_a}{2}-i \left(\Delta+\frac{\text{K}_{aa}}{\kappa_a}{|1+\alpha|}^2{|a^{in}|}^2+2\frac{\text{K}_{ab}}{\kappa_b}{|\beta|}^2 {|a^{in}|}^2+8\text{K}_{ac}\langle c^\dagger c\rangle\right)\right] \frac{\left(1+\alpha\right)}{\sqrt{\kappa_a}}=-i\frac{g\langle c \rangle}{\sqrt{\kappa_b}} \beta + \sqrt{\kappa_a} 
\end{equation}
\begin{equation}
\left[\frac{\kappa_b}{2}-i \left(-\Delta+\epsilon+\frac{\text{K}_{bb}}{\kappa_b}{|\beta|}^2{|a^{in}|}^2+2\frac{\text{K}_{ab}}{\kappa_a}{|1+\alpha|}^2 {|a^{in}|}^2+8\text{K}_{bc}\langle c^\dagger c\rangle\right)\right] \frac{\beta^*}{\sqrt{\kappa_b}}=-i\frac{g\langle c \rangle}{\sqrt{\kappa_a}}\left(1+\alpha^*\right).
\end{equation}
We note that these equations can no longer be expressed in terms of $\alpha$ and $\beta$ alone, and the explicit dependence on $a^{in}$ will result in gain saturation effects absent from the third-order stiff pump expressions.  The unsaturated reflection gain for sufficiently small signal power can be calculated from theses two equations by assuming $a^{in}=0$ as
\begin{equation}
{|\alpha|}^2=\frac
{16g^4 {n_p}^2+8g^2{n_p}\left(-4\Delta_{m}\Delta_{n}+{\kappa_a}{\kappa_b}\right)+\left(4{\Delta_{m}}^2+{\kappa_a}^2\right)\left(4{\Delta_{n}}^2+{\kappa_b}^2\right)}
{16g^4 {n_p}^2+8g^2{n_p}\left(-4\Delta_{m}\Delta_{n}-{\kappa_a}{\kappa_b}\right)+\left(4{\Delta_{m}}^2+{\kappa_a}^2\right)\left(4{\Delta_{n}}^2+{\kappa_b}^2\right)}
\label{eq:G4}
\end{equation}
where $\Delta_{m}=8\text{K}_{ac}n_p+\Delta$ and $\Delta_{n}=8\text{K}_{bc}n_p-\Delta+\epsilon$.

Again substituting $\kappa_{a,b}=\kappa \mp \Delta\kappa/2$, we can solve  $\Delta_{\text{max}\;\text{G}}$ to first order in $\epsilon$ and $\Delta \kappa$ and find
\begin{equation}
\Delta_{\tiny{\textrm{max G}}} \simeq 
\left(K_{bc} - K_{ac}\right)n_p + \frac{\left(K_{bc} + K_{ac}\right)}{ x_{-}}\frac{\Delta \kappa}{2}\kappa n_p 
+\left(\frac{1}{2}-\frac{\kappa \Delta \kappa}{2 {x_{-}}^2}x_{+}\right)\epsilon
\end{equation}
where $x_{\mp}=4g^2n_p\mp 4\left(K_{bc}+K_{ac}\right)^2{n_p}^2+{\kappa}^2$.  In this fourth order expression there is a net pump-dependent frequency shift except in the special case where the pump-dependent cross-Kerr terms $K_{ac,bc}$ are equal and resonator bandwidths $\kappa_{a,b}$ are equal.
\begin{figure*}[h!]
	\includegraphics{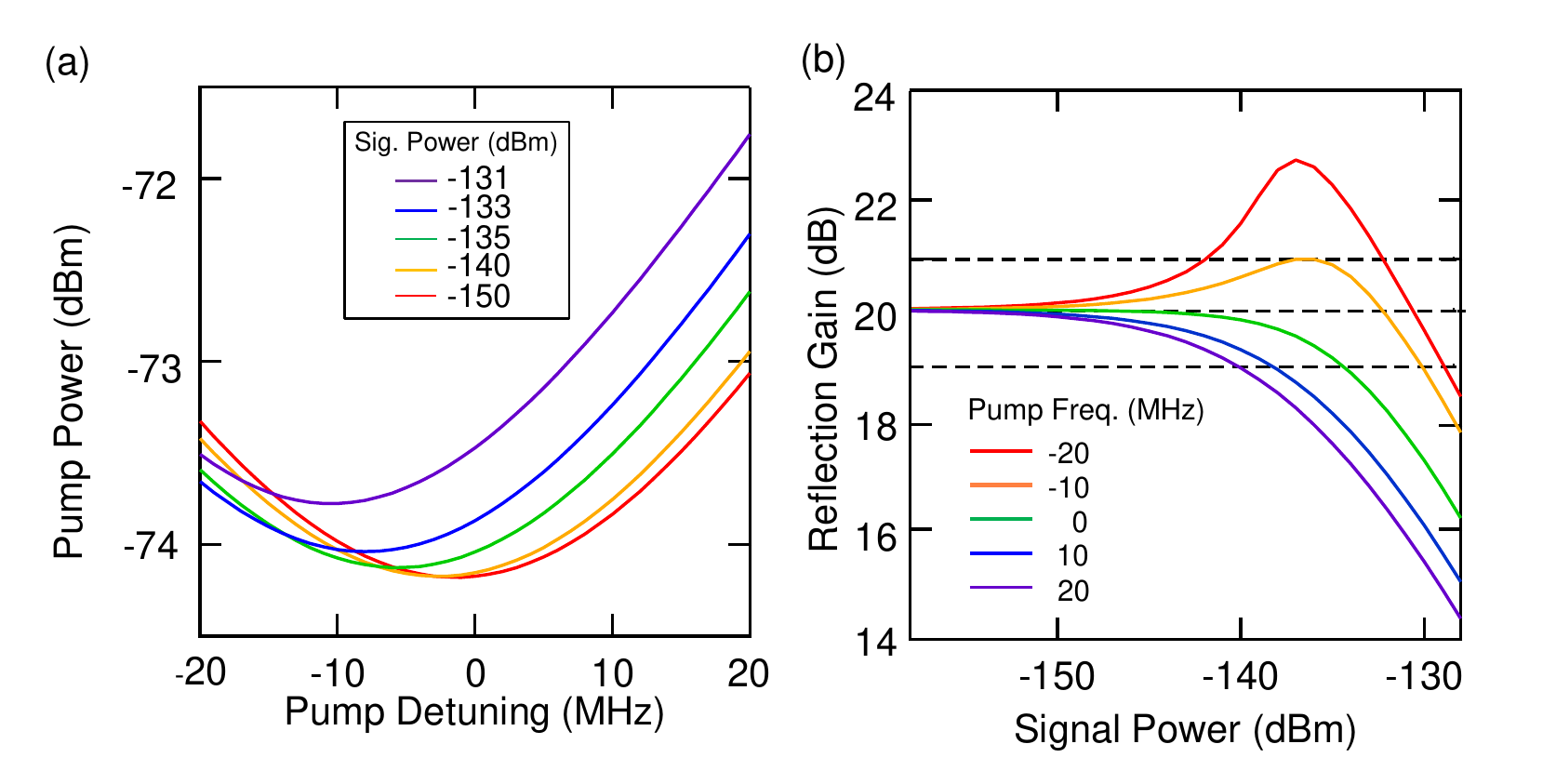}
	\caption{ \textbf{(a)} Family of 20~dB gain with different signal power. 20~dB curve moves toward lower frequency when signal power increases due to Kerr terms. Based on equation (17) and (18), Kerr terms play an important role to affect 20 dB gain. \textbf{(b)} Saturation of gain curves with different detuning. From this figure, optimal points can be defined in two ways. Max dynamic range happens when pump detuning is $-6\;\text{MHz}$ while max flatness appears at $-1\;\text{MHz}$ pump detuning.}
\end{figure*}

\subsection{Parameter Estimation}
To most accurately obtain the coefficients of the three-wave mixing and Kerr terms of our device, we numerically calculated the Hamiltonian of the full 8-junction system.  We used measured room-temperature resistances to estimate the junction critical currents and  including estimates for the stray linear inductances produced by superconducting lines connecting each junctions, together with the effective inductance and capacitance of the $41 \Omega$ characteristic impedance resonators.  We note, however, that the primary discrepancies between data and experiment for this device derive not from the properties of the JRM itself.  Instead, they are due to the strongly-varying, frequency-dependent shifts in the mode lifetimes caused by the imperfect external impedance which oscillates as a function of frequency simultaneously for the several modes.  This behavior is not included in our model, which assumes a fixed mode lifetime for all frequencies.

\subsection{Large signal response and saturation curves}
To find the 20~dB gain points in Fig.~S2a for each input signal strength $P_{sig}=n_a \hbar \omega_a \kappa_a$ ($n_a=|a^{in}|^2$), we  impose the condition $G=|\alpha|^2=100$ and solve for $n_p$ for each value of pump detuning $\epsilon$. As is typical for parametric amplifiers (see Eq.~\ref{eq:Gsimp}), there are two solutions for $n_p$($P_p$): one above the critical pump power and one below it. We exclusively choose the lower solution, and solve for the $\Delta_{\tiny{\textrm{max G}}}$ at which the peak gain occurs. 
To find saturation curves as depicted in Fig.~3a and supplementary Fig.~S2b, we choose small signal bias conditions identified in the previous section for each pump detuning $\epsilon$ and solve the Langiven equations for the complex $\alpha$ and $\beta$ response at the small-signal max gain frequency $\Delta_{\tiny{\textrm{max G}}}$ for a succession of values for $n_a$.  For theory plots in supplementary Fig.~S3 a similar procedure is followed, however we additionally redefine the phase of $\beta$ to be zero at small signal amplitudes to match experimental data for which we are only sensitive to changes in phase vs. signal power, not the absolute value\cite{Frattini2017, Hatridge2013}.

\begin{figure*}[h!]
	\includegraphics{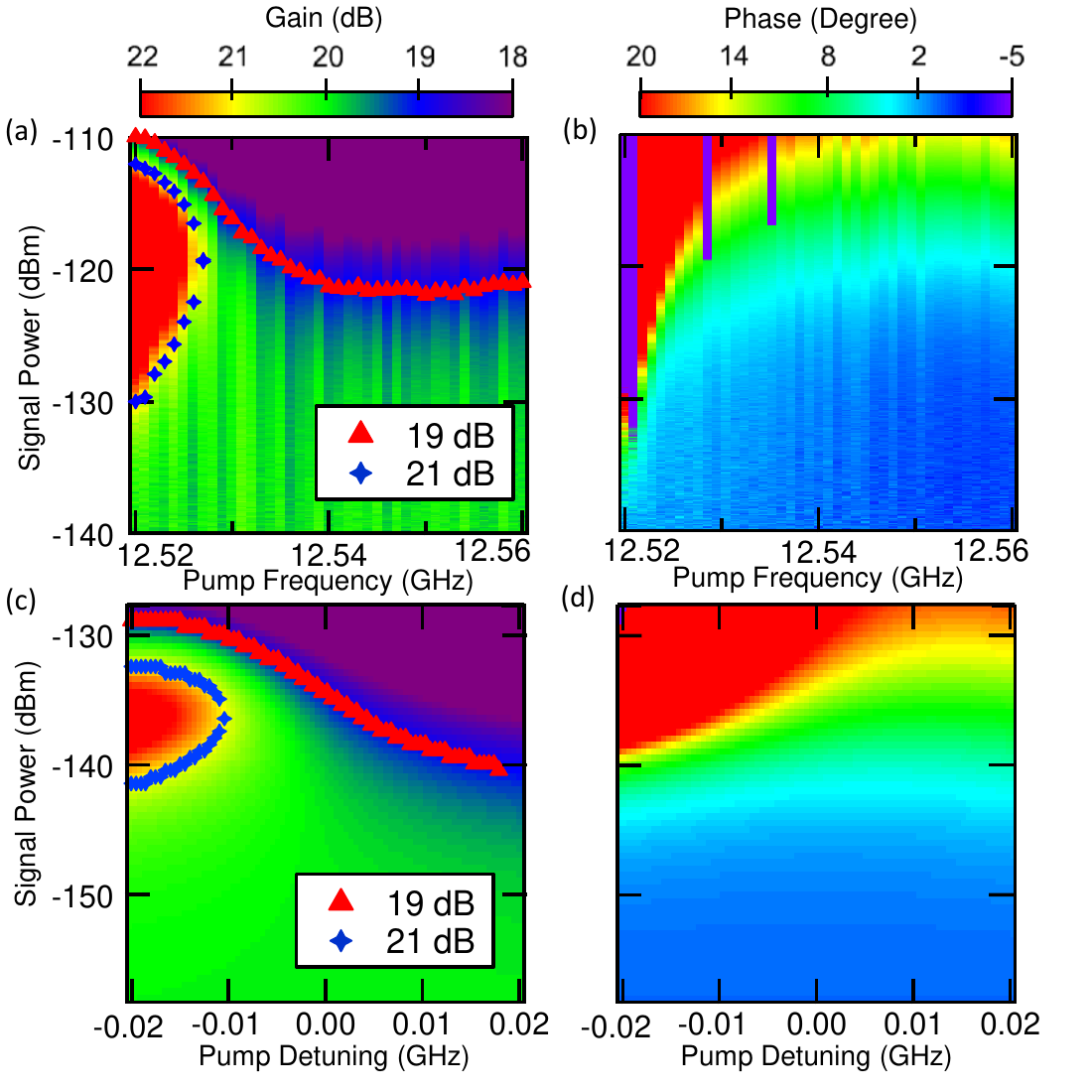}
	\caption{ \textbf{(a and c)} Experimental transmission gain amplitude (a) and relative phase (c) data vs. signal power and pump frequency of the JPC measured under the same condition as Fig.~3b. The red triangles represents the 19 dB gain points at each pump frequency, and the blue diamonds represents the 21 dB points.  Since the input and output signals are at different frequencies, a separate mixer at the difference frequency is used to convert the signals back to the same frequency so they can be compared in the VNA.  To calibrate for the unknown phase response vs. frequency of the measurement lines, the small-signal phase for each pump frequency is defined to be zero and the plotted data represent shifts relative to this value.\textbf{(b and d)} Calculated transmission gain amplitude (b) and relative phase (d) data vs. signal power and pump frequency under the same condition as Fig.~3c.  Again, the phase shifts are defined to be zero for small signal powers and the relative phase is plotted to correspond to the experimental data.}
\end{figure*}

\bibliography{bibwkmsmtvol2a}% Produces the bibliography via BibTeX.